%% file: main.tex
\documentclass[sigconf,anonymous,review]{llncs}
\usepackage{graphicx}
\usepackage{amssymb}
\usepackage{enumitem}
\usepackage{url}
\usepackage{cite}
\setlist{nosep}

\long\def\comment#1{}
\long\def\comments#1{}
\begin{document}

\title{LSTM-based Selective Dense Text Retrieval Guided by  Sparse Lexical Retrieval}

\author{
Yingrui Yang\inst{1}\orcidID{0000-0001-6454-5796} \and
Parker Carlson\inst{2}\orcidID{0000-0003-1856-5088} \and \\
Yifan Qiao\inst{3}\orcidID{0000-0001-5717-2637} \and
Wentai Xie\inst{2}\orcidID{0009-0007-7870-3100} \and \\
Shanxiu He\inst{2}\orcidID{0009-0008-8581-6733} \and
Tao Yang\inst{2}\orcidID{0000-0003-1902-3387} 
}
\authorrunning{Y. Yang et al.}
%

\institute{
Coursera Inc., USA,
\email{yingruiyang@ucsb.edu}
\and
University of California at Santa Barbara, USA, 
\email{\{parker\_carlson,  wentaixie, shanxiuhe,tyang\}@ucsb.edu}
\and
Apple Inc.,  USA,
\email{yifan\_qiao@apple.com}
}

\maketitle



\begin{abstract}

\input{ecirabs}

\end{abstract}



\input{cikmintro_ecir}
\input{ecirproblem}
\input{cikmmethod_ecir}

\input{ecirexp}

\input{cikmexp_ablation_ecir}

\input{cikmexp_beir_ecir}

\input{cikmexp_io_ecir}
\input{cikmexp_llama_ecir}

\input{cikmexp_sparse_ecir}
\input{cikmexp_compress_ecir}

\input{cikmexp_feature_ecir}
\input{ecirconcl}

\newpage
\bibliographystyle{splncs04}
\bibliography{bib/ranking.bib,bib/distill.bib,bib/2024ref.bib}
\end{document}

%% file: ecirabs.tex
\comment{
{\bf Old abstract:}
Recent studies show that neural sparse retrieval for searching  text documents with a learned neural representation
can be greatly accelerated but there are  tradeoffs  in its relevance effectiveness.
This paper proposes a relevance augmentation scheme of fast sparse  document retrieval with cluster-based selective dense retrieval called CluSD
on  a CPU-only platform.
CluSD takes an advantage of clustering under an advanced  dense retrieval model and
detects clusters of neural embeddings  that can boost sparse retrieval with  limited  extra memory space overhead and embedding selection and computing time. 
This paper provides a detailed evaluation of CluSD for searching  MS MARCO and BEIR  datasets, 
and demonstrates its effectiveness in improving the relevance metric of sparse retrieval by scoring  selected dense embedding clusters 
at a low time and space cost.

{\bf New abstract:}
 both of which have benefited from pretrained language models.
}
This paper studies fast fusion of dense retrieval and sparse lexical retrieval,
and proposes a cluster-based selective dense retrieval method called CluSD guided  by  sparse lexical retrieval. 
CluSD takes a lightweight cluster-based approach and exploits the overlap of sparse retrieval results and embedding clusters
in a two-stage selection process with an LSTM model to quickly identify relevant clusters
  while incurring limited  extra memory space overhead. 
CluSD  triggers partial dense retrieval and  performs cluster-based block disk I/O if needed.
This paper evaluates CluSD and compares it with several baselines for searching in-memory and on-disk MS MARCO and BEIR datasets. 

%% file: cikmintro_ecir.tex
\section{Introduction and Related Work}

\label{sect:intro}
Dense and sparse retrievers are two main categories of retrieval techniques for text  document search.
As shown in previous literature~\cite{Lin2021unicoil,2022LinearInterpolationJimLin, kuzi2020leveraging}, combining  sparse and dense retrieval scores with linear interpolation
can boost search relevance. 
It is important to optimize for efficiency when combining the above two retrieval approaches on a low-cost CPU-only platform.
This  is desirable in a large-scale  search system  which employs  a multi-stage search architecture, and 
runs partitioned first-stage retrieval in parallel  on a massive number of inexpensive CPU-only
machines. 
It is also critical for search on personal devices such as phones with limited computing/memory resources or battery use constraints.

Dense retrieval can be accelerated   with  approximate nearest neighbor (ANN) search using 
partial IVF cluster search~\cite{johnson2019billion, 2021Facebook-DrBoost-Lewis} or proximity-graph-based navigation (e.g. HNSW)~\cite{2020TPAMI-HNSW}.
However, there is a  significant
relevance tradeoff for a  reduced search cost in these efforts. For example,
RetroMAE~\cite{Liu2022RetroMAE}, a state-of-the-art  dense retriever,  is slow on CPU without compression, the use of
5\% IVF cluster search reduces CPU time substantially,  but there is a 7.5\% drop in MRR@10.
The use  of  OPQ quantization~\cite{Ge_2013_CVPR}, implemented in FAISS~\cite{johnson2019billion},
 further reduces memory space  to 1.2GB but causes an extra 4.6\% MRR@10 drop.
Notice that the idea of cluster-based retrieval was  explored for traditional information retrieval and 
selective search~\cite{ Can2004EfficiencyAE, 2008ACMTrans-Altingovde,2017ECIR-Hafizoglu}.

\comments{
\begin{table}[htbp]
        \centering
        \caption{Latency, relevance, and space  tradeoff  in a dense model for passage retrieval with MS MARCO Dev set}
         \resizebox{0.65\columnwidth}{!}{ 
                \begin{tabular}{ r | r  rr}
                        \hline\hline
Dense methods  & Latency & Memory & MRR@10\\ 
\hline
RetroMAE/Uncompressed & 1677ms & 27.2GB & 0.416 \\ 
RetroMAE/IVF 5\% & 107ms  & 27.2GB & 0.387 \\ 
RetroMAE /IVF 5\% OPQ  & 38ms & 1.2GB  & 0.370 \\ 
                        \hline\hline
\end{tabular}
}

 \vspace*{-6mm}
\label{tab:costbudget}
\end{table}

}


LADR~\cite{2023SIGIR-LADR}, as a follow-up study of  GAR~\cite{2022CIKM-MacAvaneyGraphReRank},
has investigated partial dense retrieval by  
using the results of a sparse retriever as a seed to select embeddings
based on a document-to-document proximity graph. This strategy  follows  the previous graph-based ANN approaches~\cite{2020TPAMI-HNSW,2021KDDgraphANN}. 
Using  sparse retrieval results as a seed  allows quick narrowing of  the search scope at a low CPU cost. 
While LADR and HNSW have demonstrated their high efficiency  to search within a time budget,
the use  of a document-wise similarity graph adds a significant online space requirement  on a low-cost computing platform. 
For example, the proximity graph  can take an extra 4.3GB of memory space~\cite{2023SIGIR-LADR} for 
8.8M  MS MARCO passages. 

Another limitation of LADR and  HNSW is the assumption of in-memory access to proximity graphs and dense vectors. 
For large datasets or higher-dimension embedding vectors, some or all of the embeddings and proximity graphs will have to 
be stored on disk, especially when such  applications desire uncompressed embeddings for  better  relevance. 
For example, the embedding dimension of the recent RepLLaMA dense retriever~\cite{ma2023finetuning} is 4096,
based on LLaMA-2~\cite{touvron2023llama}, which leads to 145GB storage space for MS MARCO passages.
It is possible that more advanced dense models may be developed to take advantages of  
large language models in the future with  a higher  embedding dimensionality (e.g. up to 12,288~\cite{NEURIPS2020_GPT3}) for 
better relevance. 
For large on-disk search, random access of dense embeddings and/or a document-level proximity graph can incur substantial 
fine-grained I/O overhead.
DiskANN~\cite{NEURIPS2019_DiskANN} and SPANN~\cite{chen2021spann} are
two on-disk ANN search algorithms for general data applications which do not  leverage sparse retrieval.
They have not been studied in the context of text document retrieval
where the importance of semantic text matching presents a unique challenge and requires new design considerations.

One recently published contemporary work, CDFS~\cite{2024SIGIR-CDFS-Yang}, addresses the above problem. 
One weakness of CDFS is that it uses probabilistic thresholding to select  dense clusters  based on a strong assumption that 
the order statistics of query document ranking is independently and identically distributed. 
This assumption is only true when the query document similarity scores for relevant and irrelevant documents follow the same distribution, 
which is rare because they are usually distributed differently.

This paper proposes a lightweight approach called CluSD 
(Cluster-based Selective Dense retrieval) guided by sparse retrieval results. 
Unlike CDFS, CluSD does not follow any statistical distribution assumption. 
CluSD limits query-document similarity computations through a two-stage cluster selection algorithm via an LSTM model, which  
exploits inter-cluster distances and the overlapping degree between top sparse retrieval results and dense clusters.
When dense clusters are not available in memory, CluSD selects and loads a limited number of clusters with efficient disk block I/O.
Thus, it can handle a large dataset with high-dimension embeddings that cannot fit in memory, with less I/O overhead compared to a
graph-based ANN approach.  

%% file: ecirproblem.tex
\vspace*{-1em}
\section{Selective Dense Cluster Retrieval}
\label{sect:background}

\comments{
\begin{table}[h]
	\centering
		\begin{small}
		\begin{tabular}{l l}
			\hline
			Symbol & Def. \\
                \hline
                $Q$ & A query \\
                $d_i$ & A document \\
                $\mathcal{D}^{+}$, $\mathcal{D}^{-}$ & Positive/negative document subsets corresponding to $Q$  \\
                $|.|$ &  The size of a set\\
		$\Theta$& Parameters of a scoring model\\
		$S(Q,d_i, \Theta)$ & The rank score of a document for a query using a model\\
                $p_i$ & Teacher's top one probability of document $d_i$,\\
			& standing for  $P(d_i|Q, \mathcal{D}^+, \mathcal{D}^-, \Theta$) \\
                $q_i$ & Student's top one probability of document $d_i$\\
                $\pi(.)$ & The rank position of a document for a query\\
                $\beta_i$ & Exponent weight bias for negative  document $d_i$\\
                $\gamma, \alpha$ & CKL hyper-parameters\\
                \hline
		\end{tabular}
		\end{small}
	\caption{Table of Symbols}

	\label{tab:data}
\end{table}
}

\textbf{Problem definition.}
Given query $q$ for searching a collection of $D$ text documents:   $ \{d_i\}^D_{i=1}$, 
retrieval  obtains the top $k$ relevant documents from this collection based on
the similarity between query $q$ and document  $d_i$ using  their
representation vectors.
With a lexical representation, each  representation vector of a document or a query is sparse.
Sparse retrieval uses  the dot product of their lexical representation vectors as the similarity function
and implements its search efficiently using an inverted index: 
$L(q) \cdot  L(d_i)$
where $L(.)$ is a lexical representation   which is a vector of weighted term tokens for a document or a query.
For BM25-based lexical representation~\cite{Robertson2009BM25}, 
$L(.)$ treats each document or a query as a bag of terms.
Term weights are scaled based on the frequency of terms within a document and across all documents in the collection.
For a learned sparse representation~\cite{Dai2020deepct, Mallia2021deepimpact, Lin2021unicoil,2021NAACL-Gao-COIL}, 
a document or a query is encoded using  a trained neural ranking model, which
produces modified representations of document or query containing both original and expanded terms.
The encoding functions for a query and document can be different. Without the loss of generality, we assume they are the same in this paper.
With a single-vector dense  representation, each  representation vector of a document or a query is a dense vector. 
Dense retrieval computes the following rank score: 
$R(q) \cdot  R(d_i)$
where $R(.)$ is a dense representation vector  of a fixed size~\cite{Karpukhin2020DPR}.

\comments{
For each document, the SPLADE term weighting revises the SparTerm model~\cite{Bai2020SparTerm} to predict the weight 
score $w_j$ of $j$-th token term in BERT WordPiece vocabulary.
\begin{equation}
\label{eq:splade}
w_j = \sum_{i \in d}  log (1 + ReLU ( H(h_i)^T E_j + b_j)) 
\end{equation}
where document $d$ consists of a sequence of BERT last layer's embeddings $(h_1, h_2, $
$\cdots, h_n)$.
$E_j$ is the BERT input embedding of the $j$-th token and $b_j$ is a token level bias.
$H(.)$ is a linear layer with activation and layer normalization. 

{\bf Combining the results of a dense retriever and a sparse retriever.}
}
Following the work of~\cite{kuzi2020leveraging, 2022LinearInterpolationJimLin, Gao2020ComplementingLR}, 
linear interpolation is used  to  ensemble  dense and sparse retrieval scores. 
The fused rank score for document $d_i$ in responding  query $q$ is:
$\alpha L(q) \cdot  L(d_i) +
(1-\alpha) R(q) \cdot  R(d_i) 
$
where $\alpha$ is a coefficient in a range between 0 and 1.
Our goal is to maintain a relevance competitive to full dense retrieval, while minimizing the time and memory overhead of cluster-based dense retrieval,
guided by sparse retrieval results.
\comments{
The ensembling method used in our evaluation is 
the reciprocal rank fusion (RRF) following Cormack et al.~\cite{Cormack2009RRF}, considering ranking positions of each candidate 
generated by a retriever and by a re-ranker. 
We use RRF
instead of linear interpolation as it has a similar performance with a simplified combination of two scoring models. 
\begin{equation}
     RRF(Q,d) =  
\frac{1}{c_1 + \pi_1(d)} + 
\frac{1}{c_{2} + \pi_2(d)},
     \label{eq:score}
\end{equation}
where $c_{1}$ and $c_{2}$ are fixed smoothing constants. $\pi_2(d)$ is the rank position based  a re-ranker formula
while $\pi_1(d)$ is the rank position based on a retriever.
}

\label{sec:clusd}

{\bf Design considerations.}
We aim to identify the opportunity to skip a substantial portion of dense retrieval
by taking the IVF clustering approach~\cite{johnson2019billion} that  groups similar documents.
This  helps CluSD to recognize similar relevant documents during inference while avoiding the
need of a document-level proximity graph. The number of clusters needs to be as large as possible (meaning a small size per cluster) to avoid unnecessary I/O and computing cost. That imposes a significant challenge for accurate and fast selection of dense clusters.
To address this  our work proposes a two-stage LSTM-based 
selection algorithm that exploits the overlap of sparse retrieval results 
with the potential dense clusters.

%% file: cikmmethod_ecir.tex
\subsection{Steps of online inference in CluSD}
\label{sect:online}

\comments{
\begin{figure}[htbp]
    \centering
    \includegraphics[scale=0.79]{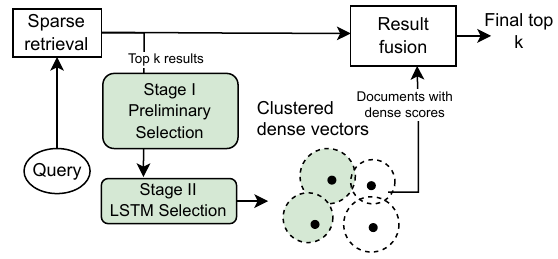}
 \vspace*{-2mm}
    \caption{The flow of cluster-based selective dense retrieval}  
 \vspace*{-3mm}
    \label{fig:flow}
\end{figure}
}


Online inference of CluSD has the following steps:
\textbf{Step 1}:   Conduct sparse retrieval to obtain top-$k$ results.
\textbf{Step 2}:  Conduct a two-stage selection process. 
Stage I selects the top-$n$ clusters from  the given $N$ dense embedding clusters, discussed in detail in Section~\ref{sect:stage1}.
Stage II  applies an LSTM model to the $n$ top clusters
to choose  a limited number of dense clusters to evaluate, discussed in Section~\ref{sect:lstm}. 
\textbf{Step 3}: Expand the top-$k$ sparse retrieval results to include the documents vectors of these dense clusters and
fuse their sparse and dense scores with linear interpolation.  

\comments{
Notice that  the two ranked lists from two retrievers are usually not fully overlapped and  some documents only exist in one list. While
different  treatments on these documents affect fusion performance,
we resort  to imputing the score of such a document using  the top $k+1$ score   in the missed rank list scaled by 
a factor.
For example,  we  impute the lexical scores of new documents added  by dense retrieval
using  the scaled lexical score of  top $(k+1)$ in the sparse retrieval list. We use scale $0.95$  in our evaluation.
}


{\bf Time  and space  cost for CluSD.} 
The embedding space is $O(D)$  for a corpus with $D$ vectors.
For MS MARCO, this  cost is around 27GB without compression. Quantization can reduce the total size to about  0.5GB  to 1GB, depending on codebook parameters.
Then this cost becomes a fraction of sparse retrieval index cost.  
Excluding embeddings, the extra space overhead for CluSD is to maintain  the similarity among centroids of $N$ clusters.
We only maintain the top $m$ cluster neighbors for each cluster in terms of their centroid similarity. 
For our evaluation, $m=128$ and this reduces the extra space overhead as $O(N)$, which
is negligible compared  to the embedding space with $N << D$. 
For example,  with $N=8192$,  the inter-cluster graph for MS MARCO passages takes about 5MB with quantization.
Excluding sparse retrieval cost, the time  cost of 
CluSD is dominated by the cost of cluster selection and  dense similarity computation of selected dense clusters.
The former is about $O(n)$ while the latter is
proportional to the number of clusters selected, which is capped by $O(\frac{D}{N} n)$. 
In  our evaluation of MS MARCO and BEIR, $n$ is chosen to be 32, and
the actual number of clusters selected and visited is about 22.3.
Thus, the overall latency time overhead introduced by CluSD is fairly small.

\subsection{ Stage I of Step 2: Selection of  top $n$ candidate clusters} 
\label{sect:stage1}
\comments{
As discussed earlier,  we control 
the input length  of the above  LSTM model  to be  $n$  as 
the number of top candidates clusters to be examined, and then the inference 
}

Given  $N$ clusters grouped using dense document embeddings, 
we devise a strategy to conduct a preliminary selection that
prunes low-scoring clusters that contain  only less relevant documents.
A two-stage approach is motivated by the fact that it is too time consuming for an LSTM model, used in the second stage, to examine all clusters.
Thus, we choose the top-$n$ candidate clusters first.
\comments{
There are  $n$ nodes corresponding to these clusters in the LSTM model.
We will discuss how to truncate with a prioritization shortly below.
}
\comments{
Table ~\ref{tab:LSTMclustertime} shows that the number of clusters to be examined
varies from 32 to 512, the LSTM inference time increases from  2.8ms to 33.9ms.
As we would like to make a quick decision given the overall  time budget,
we devise another strategy to prune clusters that may not provide a meaningful boosting
of relevance quality. 


\begin{table}[h]
	\centering
		\resizebox{0.68\columnwidth}{!}{ 
		\begin{tabular}{r | rrrrr}
			\hline\hline
			  \#Clusters examined & 512 & 256 & 128 & 64 & 32 \\
\hline
            Time (ms) & 33.9 & 16.2 & 6.2 & 3.8 & 2.8\\
                \hline\hline
		\end{tabular}
	}
	\caption{LSTM inference time with different input lengths.}
	\label{tab:LSTMclustertime}
\end{table}

\comments{
\begin{figure}[htbp]

    \includegraphics[scale=0.27]{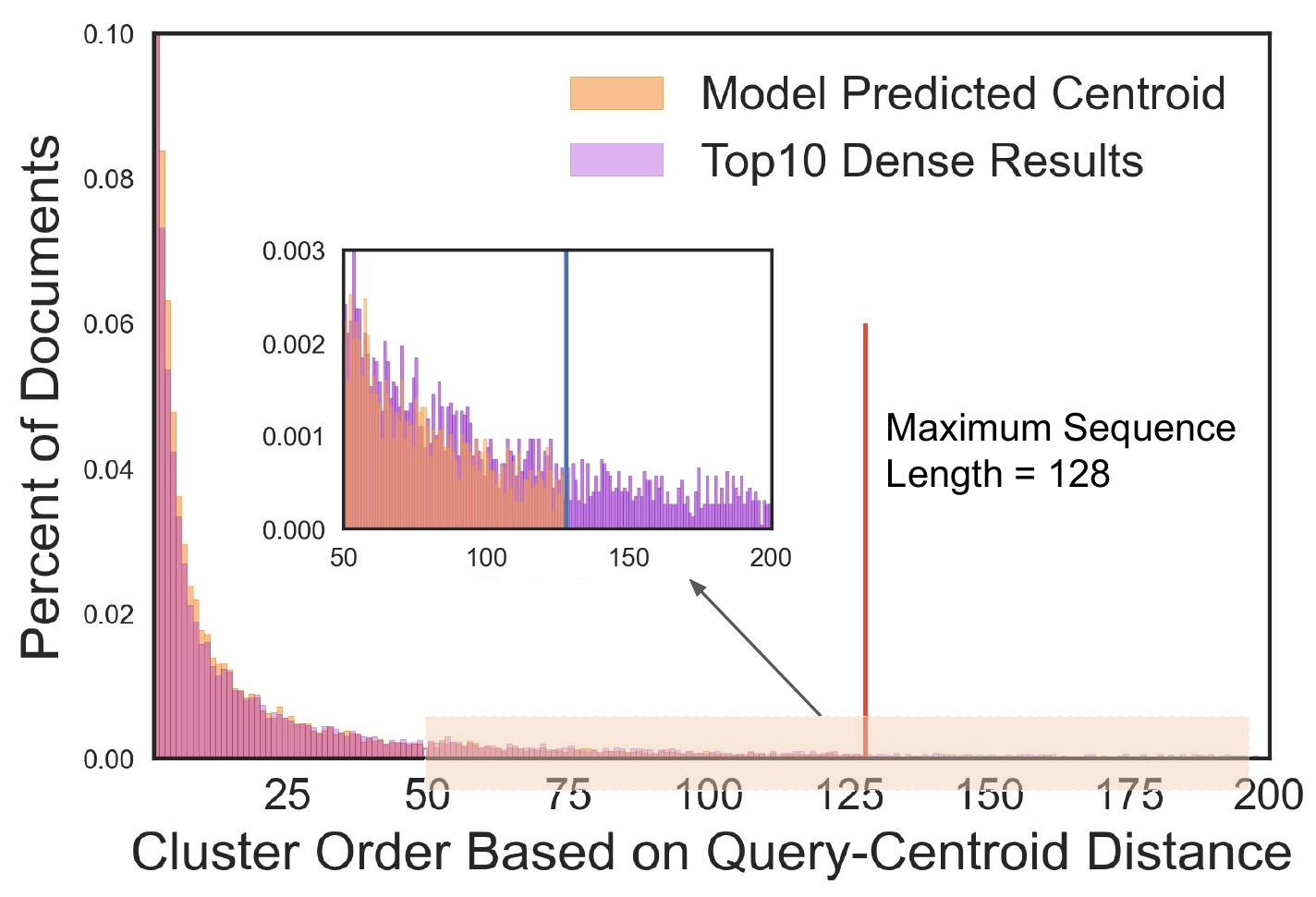}
    \caption{Distribution of Good Documents }
    \label{fig:dist}
\end{figure}
}
}

To prioritize clusters, one idea is to adopt their query-centroid distance used
in IVF cluster selection.  We find this does not work well for CluSD.
For example, if we rank clusters based on the query-centroid distance for MS MARCO passages, 
around 10\% of top-10 dense results reside in clusters ranked beyond the 175th cluster, which means we would have to 
select 175 clusters in order to recover 90\% of top-10 documents. Visiting all of these clusters is still too slow
and 
CluSD only visits about 22.3 clusters on average for MS MARCO Dev set.

Our method   for the preliminary selection at this stage is to rely on
the degree of overlap between dense clusters and the top sparse retrieval results. 
Specifically, we divide $k$ top sparse retrieval results into $v$ top sparse result bins denoted $B_1, \cdots, B_v$.  
We  assign each embedding cluster $C_i$ with a priority vector $(P(C_i, B_1), \cdots, P(C_i, B_v))$ where
$P(C_i, B_j)=  |C_i  \cap  B_j|$,
which is the number of documents in top sparse bin $B_j$ included in cluster $C_i$.
A larger $P(C_i, B_j)$ value indicates a higher priority for $C_i$ with respect to sparse result position bin $B_j$.
In our evaluation with  retrieval depth $k = 1000$ for the MS MARCO and BEIR datasets,
we use $v=6$ result bins, divided into the top-10 results, top 11-25, top 26-50, top 51-100, top 101-200, top 201-500, and top 501 to $k$ documents.
The main reason to divide sparse results into bins is to reduce the time complexity of LSTM for cluster selection with 
a shorter sequence. The reason to have a non-uniform bin size is that the importance of rank positions derived by sparse 
retrieval is nonlinear. For example, the top 1 to 50 results are much more important than the results 
ranked from 50 to 1,000, requiring finer grained partitioning.
Then, we perform multikey sorting based on cluster priority vectors $(P(C_i, B_1)$, $\cdots$, $P(C_i, B_v))$. The clusters are sorted first based on $(P(C_i, B_1)$, then $(P(C_i, B_2)$ if there's a tie. 
If there is a tie between two clusters, then we rank them by their centroid distance to the query.
The top-$n$ ranked candidate clusters will be  the  input for the LSTM model described below.

\comments{
Section~\ref{sect:evalLSTM} studies the effectiveness of features used in Stage II, and explains our choice of features for preliminary cluster selection. 
For the choice of $n$,
our evaluation finds  that a relatively small number of candidate clusters, such as $n=32$, yields competitive results  for our
evaluations when $k=1000$.

}
\subsection{Stage II of Step 2: Selection with LSTM}
\label{sect:lstm}

 \vspace*{-9mm}

\begin{figure}[htbp]
    \centering
    \includegraphics[scale=0.8]{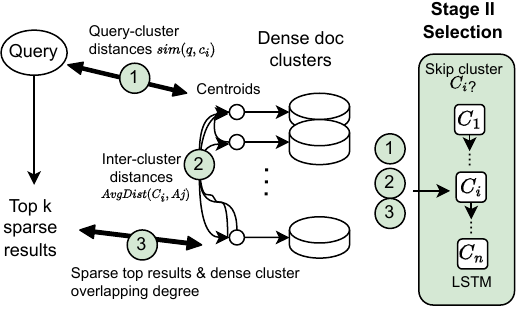}
    \caption{Illustration of CluSD and its features.}
    \label{fig:features}
\end{figure}
 \vspace*{-5mm}

\comments{
The LSTM model outputs 1 for each cluster node if this cluster should be visited to perform the corresponding dense retrieval.
The goal of this procedure  is to visit  dense clusters as few as possible while still delivering a competitive fusion performance.
The number of predicted clusters are different across queries and the overall budget is controlled by the prediction threshold.
}

To select which dense clusters we visit, we adopt a simple LSTM model~\cite{hochreiter1997long}
which takes a sequence of clusters as input along with their features. We choose an LSTM because it can 
capture cluster selection order as a sequence. Selected clusters can influence what will be selected next based on 
their semantic distance and other features. This allows CluSD to select clusters accurately to 
minimize I/O and computing cost.
The time complexity of the LSTM model is limited by $O(n)$,
where the $n$ top clusters are  given from  Stage I.

The LSTM sequentially visits each cluster, denoted $C_1$ to $C_{n}$. 
For each cluster $C_i$, the LSTM predicts a score $f(C_i)$ such that $0\leq f(C_i) \leq 1$.
If $f(C_i) \ge \Theta$, where $\Theta$ is a set prediction threshold, then cluster  $C_i$ should be visited. 
$\Theta$ controls the tradeoff between efficiency and relevance; in practice, $\Theta$ can be tuned based on the overall latency requirement.
In our experiments,  the default setting of
CluSD is to use a threshold of 0.02,
yielding 22.3 clusters selected on average. Due to the page limit, we have omitted a study of
 the impact of varying this threshold on the number of clusters selected.

\comments{
When the dense vectors of all documents are grouped into $N$ clusters, 
we limit our cluster search scope to top $n$ clusters in terms of query and cluster centroid similarities. 
we limit $n=128$ in our evaluation out of 1000 clusters.
}

As depicted in Figure~\ref{fig:features}, the feature input vector for the current cluster  $C_i$ for the above LSTM computation to produce $f(C_i)$
is composed of the following three groups of features:

\comments{
\[ sim(q,c_i),   
\{ AvgDist(sim(c_i, c_l) \mbox{ for } c_l \in A_j) \}^u_{j=1}, 
\]
\[ 
     \{  |c_i  \cap  B_j|    \}^{v}_{j=1},  
\{  AvgScore (c_i  \cap  B_j)    \}^{v}_{j=1}. 
\] 
}
\begin{itemize}[leftmargin=*]
\item \textbf{Query-cluster similarity:} $sim(q,c_i)$.
The similarity distance of this query $q$ with the centroid $c_i$ of cluster $C_i$.  


\item \textbf{Inter-cluster similarity:} $AvgDist(C_i, Aj)$ for $1 \leq j \le u$.
Given the $n$ sorted clusters derived from Stage I, we uniformly partition  these $n$ consecutive  clusters into $u$ consecutive cluster bins:
$\{ A_1, A_2, \cdots, A_u\}$. Then we define
\[
AvgDist(C_i, Aj) = \frac{1}{|A_j|} \sum_{ c_l \in A_j}
 sim(c_i, c_l) 
\]
where  $c_l$ is the centroid representing cluster $C_l$ in bin $A_j$.
The above formula computes the mean distance between centroid $c_i$ and the centroid of each cluster within  Bin $A_j$.


The purpose of this feature group is to capture the distances of a cluster to previously examined clusters and unexamined clusters 
following the LSTM flow. 
When the mean distance  between cluster $c_i$  and clusters in a previously-examined  bin is close, if  
many clusters in that bin have been selected, cluster 
$c_i$ may have a good  chance to  be selected.
In our evaluation with MS MARCO and BEIR datasets, we found that $u=6$ is appropriate.
As discussed earlier in Section~\ref{sect:online}, to reduce the extra space cost, 
we only maintain top-$m$ cluster neighbors of each cluster where $m$ is not large  (128 in our evaluation).
    

\item \textbf{Cluster overlap:} $P(C_i, B_j)$ and $Q(C_i,   B_j)$     for $1 \leq j \le v$.

This group of features captures the position-weighted and score-weighted overlap degree of this embedding cluster $C_i$ with the top sparse results.
Stage I in Section~\ref{sect:stage1} divided the top-$k$ sparse retrieval results into $v$ position bins
$B_1, \cdots B_v$ and defined   $P (C_i, B_j)$ as  the count-based overlap measure.
Now, we define a score-weighted overlap degree $Q (C_i,   B_j)$     
as the average sparse rank score of documents that  are in both $C_i$ and sparse position bin $B_j$.  
Namely,
\[
Q (C_i,   B_j) = \frac{ \sum_{d \in C_i \cap B_j} SparseRankScore(d)} { |C_i  \cap  B_j|}.    
\]

Clusters with high $Q$ scores for the top bins contain many top sparse results and may be selected.




\end{itemize}
{\bf Training of LSTM.} 
We assume that the training data  contains
a set of queries and relevance-labeled documents,  that allows the computation of a ranking metric such as MRR or NDCG.
For example, from MS MARCO training set we randomly sample 5000 training instances, where each training instance contains a query and  one or more  relevant documents. 
To train the LSTM for cluster selection, 
for each training instance, we mark the positive and negative examples for  top cluster selection.
If a cluster contains one of top-10 dense retrieval results, we mark this cluster as positive otherwise negative.
This approximately indicates that such an embedding cluster should be visited for a fusion.

%% file: ecirexp.tex
\section{Evaluations}

{\bf Datasets and measures}.
Our evaluation uses the MS MARCO dataset with 8.8 million passages~\cite{Campos2016MSMARCO,Craswell2020OverviewOT} 
for training and in-domain search,
and the  13 publicly available BEIR datasets~\cite{thakur2021beir} 
for zero-shot retrieval. 
The size of BEIR data sets ranges from 3,633 to 5.4M documents.
To report the mean latency and 99 percentile latency, 
we test queries multiple times using a single thread on  a 2.25GHz AMD EPYC 7742  CPU server with PCIe SSD.

{\bf Models and parameters}.
For sparse retrieval, we use
SPLADE~\cite{Formal2021SPLADE,Formal_etal_SIGIR2022_splade++,Lassance2022SPLADE-efficient} 
with hybrid thresholding~\cite{2023SIGIR-Qiao} and  BM25-guided traversal~\cite{mallia2022faster,qiao2023optimizing}:
1) SPLADE-HT1 with index space 3.9GB and 31.2 ms retrieval latency.
We also test on three other sparse models:
2) uniCOIL~\cite{ Lin2021unicoil,2021NAACL-Gao-COIL} with 1.2GB index and 35.5ms latency;
3) LexMAE~\cite{shen2023lexmae} with 3.7GB index and 180ms latency;
4) BM25-T5 that applies  BM25 with  DocT5Query document expansion~\cite{Cheriton2019doct5query}.  
BM25-T5 has a 1.2GB index and 9.2ms latency. 
For dense retrieval, we adopt the recent dense models
SimLM~\cite{Wang2022SimLM}, RetroMAE~\cite{Liu2022RetroMAE}, and
RepLLaMA~\cite{ma2023finetuning}. 
We use  K-means in the FAISS library~\cite{johnson2019billion} to derive dense embedding clusters.
RetroMAE-2~\cite{liu-etal-2023-retromae} is not used because its checkpoint is not released.

Our evaluation implementation uses C++ and Python.
\comments{
For sparse retrieval, we use two fast versions of SPLADE and a SPLADE-efficient model, as well as BM25-T5. 
The implementation of SPLADE models follows 2GTI optimization~\cite{qiao2023optimizing}. 
SPLADE-HT1 is the SPLADE model trained with hybrid thresholding scheme~\cite{2023SIGIR-Qiao} using $\lambda_{t}=1$, 
SPLADE-effi-HT3
is the SPLADE-efficient model~\cite{Lassance2022SPLADE-efficient} trained with hybrid thresholding scheme using $\lambda_{t}=3$. Both of these two models use the "accurate" configuration in 2GTI during retrieval, while SPLADE-HT1-fast uses the "fast" configuration.
The sparse retrieval code for Splade and BM25-T5 uses the code from PISA~\cite{mallia2019pisa}
with some optimization~\cite{Lassance2022SPLADE-efficient,2023SIGIR-Qiao}.
}
To train the CluSD model, we sample 5000 queries from the MS MARCO training set 
and the hidden dimension for LSTM is 32. The models are trained for 150 epochs. 
For sparse and dense model interpolation, we use min-max normalization to rescale the top results per query. 
For  BM25-T5, the interpolation  weights are  0.05 and 0.95 for sparse  and  dense scores, respectively. 
For other sparse retrieval models, they are  0.5 and  0.5.  
The results are marked with tag $^\dag$ when 
statistically significant drop is observed compared to the baseline marked with $\blacktriangle$ at 95\% confidence level. 


{\bf Time budget.}
We mainly use a time budget of about 50ms including the sparse retrieval time for the average latency,
and the use of this budget allows us to choose the algorithm parameters accordingly.
We choose this budget  because
in a search system running as an interactive web service,  a query with multi-stage processing
needs to be completed within a few hundred milliseconds.
A first-stage retriever operation  for a data partition needs to be completed within several tens of milliseconds without query caching.
 \vspace*{-6mm}

%% file: cikmexp_ablation_ecir.tex

\begin{table*}[htbp]

	\centering
     \caption{
    Cluster-based in-memory search  with and without space constraints
    }
\label{tab:compare} 
		\resizebox{0.85\columnwidth}{!}{
		\begin{tabular}{  r| l  |llll l|r}
			\hline\hline
			  & 
& \multicolumn{2}{c} { {\bf MSMARCO Dev} }& {\bf DL19}& {\bf DL20} & {\bf BEIR} 
& {\bf  Latency}    \\
			 &\% D
& {MRR@10}& {R@1K}& {NDCG@10}&{NDCG@10}& {NDCG@10}& ms \\
   \hline

    \multicolumn{2}{c}{}
  & \multicolumn{5}{c}{\bf{Uncompressed flat setting. Embedding space  27.2GB}}\\
     \hline
 D=RetroMAE  & 100 
& 0.416$^\dag$& 0.988& 0.720&0.703& 0.482 & 1674.1 \\
$\blacktriangle$ S $+$ D   &100 
& 0.425& 0.988& 0.740&0731& 0.520~&  1705.3 \\
  S $+$ CDFS &0.45
& 0.424& 0.987& --&-- & 0.517 & 46.0 \\
  {\bf S $+$ CluSD} &0.3
& 0.426& 0.987& 0.744&0.734 & 0.518 & 44.4 \\
              \hline
    \multicolumn{2}{c}{} &
\multicolumn{5}{c}{\bf{OPQ $m=128$. Embedding space  1.1GB}}\\
              \hline
 $\blacktriangle$  S $+$ D-OPQ &100 
& 0.416 & 0.988 & 0.737 & 0.732 & 0.515& 600.1 \\
S$+$D-IVF &10
& 0.404$^\dag$& 0.987& 0.713&0.722& 0.513 &  126.4 \\
S$+$D-IVF 
&5
& 0.394$^\dag$& 0.987& 0.687&0.706&0.507&  80.0 \\
S$+$D-IVF 
&2
& 0.374$^\dag$& 0.986 & 0.656&0.700& 0.499 & 52.5 \\
            S$+$CDFS  &0.45& 0.415 & 0.986 & 0.740& 0.730 &- & 43.3 \\
 \textbf{S $+$ CluSD}  &0.3
& 0.417& 0.986& 0.742&0.735& 0.514  & 42.6  \\
             \hline
\comments{
    \multicolumn{2}{c}{} &
             \multicolumn{6}{c}{\bf{DistillVQ $m=128$. Embedding space  1.38GB}}\\
\hline
IVF  & 2  & 0.365$^\dag$ &     0.899$^\dag$ &  --& -- & -- &  20.8ms\\
CluSD & 0.3 & 0.393 &  0.977&  --&  -- & --&  9.7ms\\
S+ IVF & 2    &0.392$^\dag$ & 0.987&--&  -- & --&   52.0ms\\
S+CluSD &0.3 &   0.417 &        0.987 & --& -- & -- & 40.9 ms\\
\hline

             \hline
            & & \multicolumn{10}{c}{\bf{OPQ $m=64$. Embedding space 0.6GB}}\\
            \hline
           $\blacktriangle$  S $+$ D-OPQ  &100 & 0.409 & 0.986 & 0.718 &  0.721  &0.501 & 0.402 & 0.986 & 0.717 &  0.719 & 0.508 & 321.6 \\
           S$+$D-IVFOPQ &10& 0.405 & 0.986 & 0.716 & 0.734 & 0.502 & 0.393$^\dag$& 0.986& 0.676&0.730& 0.505& 75.6 \\
           &5& 0.397$^\dag$& 0.986 & 0.704 & 0.733  & 0.497 & 0.384$^\dag$& 0.985& 0.659&0.717& 0.500& 55.0 \\ 
            &2& 0.383$^\dag$& 0.985 & 0.680 & 0.738 & 0.487 & 0.368$^\dag$& 0.985 & 0.643&0.707& 0.493 & 42.4 \\
            \textbf{S $+$ CluSD}  &0.3& 0.414 & 0.986 & 0.743& 0.726 & 0.511 & 0.403 & 0.987& 0.729&0.724 & 0.506 & 40.8 \\

}
              \hline\hline
		\end{tabular}

		}
   \vspace*{-5mm}
\end{table*}

{\bf Cluster-based retrieval with in-memory data}.
Table~\ref{tab:compare} compares CluSD with a few baselines for cluster-based selective retrieval 
in searching MS MARCO and BEIR datasets with and without compression.  
The dense model is RetroMAE and sparse model is SPLADE-HT1.
IVFOPQ uses a proportion of top dense clusters sorted by the query-centroid distance.
OPQ quantization from FAISS is configured with  the number of codebooks as $m=$ 128 or 64.
We report performance of IVF clustering using the top 10\%, 5\%, or 2\% clusters respectively.
The notation ``S+D'' in Table~\ref{tab:compare} (and later in other tables)
reports the fused performance of a sparse model 
with linear interpolation.
Marker  ``\%D'' means the approximate percentage of document embeddings  evaluated based on the number of clusters selected.
The last column marked with  ``Latency'' is the mean single-query time for MS MARCO Dev set.
CluSD uses setting  $N=8192$ and 
$n=32$,
which results in  selection of 22.3 clusters on average per query
and allows CluSD to meet  the latency requirement.
The results in Table~\ref{tab:compare} shows that 
under the same time budget, CluSD outperforms partial dense retrieval with IVF search in relevance.
Additionally, CluSD performs slightly better than CDFS in terms of both relevance and latency because CluSD's LSTM effectively selects less clusters to search than CDFS.
 

\comments{
Rows 6 and 11 of Table ~\ref{tab:mainspace} 
 show that selective IVF cluster search using  query-centroid distances 
can be under 130ms  when the percentage of clusters searched  drops from 100\% to 10\%. But the relevance drop is significant.  
Moreover, CluSD is as  fast as IVFOPQ  top 2\% search at m=64 (Row 14 vs. Row 13) and 2x fast at m=128 while
its relevance is much higher (Row 9 vs. Row 8).
There is no statistically significant difference between CluSD and the oracle at each compression setting, which 
indicates CluSD effectively minimizes visitation of unnecessary  dense clusters. 
}


\textbf{CluSD vs. graph navigation}. 
Table~\ref{tab:maintime} compares  CluSD with  selective dense retrieval based on proximity graph navigation
including HNSW and LADR under time budget around 50ms.
HNSW's expansion factor parameter (ef) is set as 1024. 
For LADR, we use its default setting with seed = 200, number of neighbors = 128 and use an exploration depth of 50 to meet the latency budget.
The takeaway from this table is that  relevance of CluSD is better than HNSW 
while CluSD has similar or slightly better relevance without incurring  significant  extra
space cost, compared to LADR.
When the time budget is around 25ms, the takeaway is similar.

\begin{table*}[htbp]
\vspace*{-5mm}
        \centering
        \caption{CluSD vs. graph navigation methods 
}
 \resizebox{0.7\columnwidth}{!}{
		\begin{tabular}{r |ll|l|l|rr}
			\hline\hline
			& \multicolumn{2}{|c|}{\bf{MSMARCO Dev}}& \bf{DL19}& 
\bf{DL20}& \bf{Latency}& \bf{Space}\\
			 & {MRR@10}& {R@1K}& \small{NDCG} &   \small{NDCG} & Total(ms)& GB\\
			\hline
        \multicolumn{7}{c}{\bf{Time Budget = 50 ms, S=SPLADE-HT1}}\\
        \hline
            D=SimLM & 0.411$^\dag$ & 0.985$^\dag$& 0.714 &  0.697 & 1674.1 & 27.2 \\
          S& 0.396$^\dag$& 0.980$^\dag$& 0.732   & 0.721    &   31.2 & 3.9 \\
             $\blacktriangle$  S $+$ D  & 0.424 & 0.989 & 0.740 & 0.726 & 1705.0 & 31.1 \\
              \hline
             HNSW & 0.409$^\dag$& 0.978$^\dag$& 0.669 & 0.695 & 54.4 & 40.3 \\
           S $+$ HNSW & 0.420 & 0.987 & 0.718 & 0.723 & 59.6 & 40.3 \\
            S $+$ LADR & 0.422 & 0.984$^\dag$& 0.743&0.728& 43.6& 35.4\\
            \bf S $+$ CluSD & 0.426 & 0.987 & 0.744 & 0.724 &46.3& 31.1\\
			\hline\hline
		\end{tabular}  }
	\label{tab:maintime}
 \vspace*{-5mm}
\end{table*}


\comments{
\textbf{CluSD vs. graph navigation}. 
Table~\ref{tab:maintime} compares  CluSD with  selective dense retrieval based on proximity graph navigation
including HNSW and LADR under time budget around 50ms with dense model SimLM.
For HNSW, we set its expansion factor parameter (ef) as 1024, 
For LADR, we use its default setting with seed = 200, number of neighbors = 128 and use an exploration depth of 50 to meet the latency budget.
The takeaways from this table is that 
CluSD has similar relevance as graph navigation without incurring  significant  extra
space cost, compared to HNSW and LADR.
}

%% file: cikmexp_beir_ecir.tex
\begin{table*}[h]
	\centering
 \caption{Zero-shot performance in average NDCG@10 on BEIR datasets. 
}
		\resizebox{\columnwidth}{!}{
		\begin{tabular}{r r r lr  |lllll|llllll}
			\hline\hline
                 &  & &  & &  \multicolumn{5}{c|}{SPLADE-HT1+SimLM} & \multicolumn{5}{c}{SPLADE-HT1+RetroMAE}\\
                  &  & &  & {SPLADE}  &  & &  \textbf{CluSD} & \textbf{CluSD} & \textbf{CluSD} & & &  \textbf{CDFS} & \textbf{CluSD} & \textbf{CluSD} & \textbf{CluSD}
\\
		    & \small{BM25} &{SimLM}  &  {RetroMAE} &{-HT1} & \small{$\blacktriangle$ 
 flat} &rrk&\textbf{flat}& \textbf{m=128}& \textbf{m=64}& \small{$\blacktriangle$ flat} &rrk &\textbf{flat}&  \textbf{flat}& \textbf{m=128}& \textbf{m=64}
\\
              \hline

                            \hline
                \textbf{Avg.} & 0.440 & 0.429  &  0.482&0.500 & 0.518 &0.496&  0.516 & 0.514 & 0.511 & 0.520 &0.483 &0.517& 0.518 &0.514& 0.506 \\
            \hline\hline
		\end{tabular}
		}
	
 \vspace*{-5mm}
	\label{tab:beir}

\end{table*}

\textbf{ Detailed zero-shot performance with BEIR}.
\label{sect:evalbeir}
Table~\ref{tab:beir} lists  the performance of  CluSD  
with  SimLM and  RetroMAE after SPLADE-HT1 retrieval in searching each of 13 BEIR datasets.
The LSTM model of CluSD  is trained with the  MS MARCO training set,
and it is directly applied to each BEIR  dataset to select clusters without further tuning. 
As a baseline, column ``$\blacktriangle$ flat'' is to fuse  the uncompressed full dense retrieval  with SPLADE-HT1 sparse  retrieval.
Column marked ``rrk'' is to rerank   top 1,000  sparse retrieval results interpolated with their dense scores. 
Columns with ``CluSD-flat'', ``CluSD-m=128'' and ``CluSD-m=64''  are  CluSD results  using uncompressed or compressed embeddings with
m=128 or m=64. 
This table also includes a column marked with ``CDFS-flat'' which is  the  selective fusion outcome with CDFS using uncompressed RetroMAE embeddings. 

Table~\ref{tab:beir} shows
that the relevance of CluSD with selective fusion of sparse and dense results is higher than each individual retriever,
and CluSD with limited dense retrieval   performs  closely to   oracle ``$\blacktriangle$ flat'',
and delivers good NDCG@10 under two compression settings. CluSD is competitive to CDFS for the fusion of SPLADE and RetroMAE.

%% file: cikmexp_io_ecir.tex

{\bf  CluSD vs. baselines with in-memory or on-disk data}.
Table~\ref{tab:disktime} examines the performance  of CluSD fused with SPLADE when MS MARCO passage 
embeddings cannot fit in memory and are stored on an SSD disk.
We use  $N=65,000$ for CluSD, and the extra space for quantized inter-cluster distances takes about 40MB. 
This setting here is larger than the $N=8192$ setting in Table~\ref{tab:compare}  and we choose that
because a smaller   cluster size gives a flexibility in  reducing the disk I/O size when 
the total number of clusters selected by CluSD is controlled.
We also compare other selective dense retrieval baselines using  reranking or proximity graph navigation.
The mean response time (MRT) and 99\textsuperscript{th} percentile (P99) include
the total sparse and dense retrieval time for methods noted ``S+*'', otherwise only the dense retrieval time.
The MRT cost breakdown for I/O and CPU time is also listed. 
Times reported are in milliseconds. 
``\%D'' is the approximate  percentage of document embeddings evaluated based on the number of clusters fetched from the disk.

Reranking simply fetches top-$k$ embeddings  from the disk for fusion. 
LADR is designed for in-memory search~\cite{2023SIGIR-LADR} and we run it by  assuming  memory can sufficiently host the proximity  graph while letting LADR access  embeddings from the disk. 
Excluding the I/O cost portion of LADR in this table, the in-memory search cost of LADR is marked in the ``CPU'' column.
We test two configurations of LADR, its default with 128 neighbors, 200 seed documents, and a search depth of 50,
as well as a faster configuration, selected such that its CPU time is similar to CluSD. 
LADR\textsubscript{fast} uses 128 neighbors, 50 seed documents, and a depth of 50.
DiskANN~\cite{NEURIPS2019_DiskANN} assumes the graph and original embeddings are on disk  while the memory  hosts
compressed embeddings for quick guidance.  SPANN~\cite{chen2021spann} searches disk data in a cluster-based manner based on query-centroid distances.
Both DiskANN and SPANN  are designed for on-disk search from scratch without fusing with others, and thus we simply fuse their outcome with SPLADE results.
HNSW~\cite{2020TPAMI-HNSW} is not included because its in-memory relevance  is similar as DiskANN while it is not designed to be on disk. 

 \begin{table}[htbp]
 
        \caption{Search when uncompressed embeddings are on disk for SPLADE + SimLM}
        \centering
 \resizebox{0.8\columnwidth}{!}{
                \begin{tabular}{r|c|ll|rr|rr}
                        \hline\hline
                        & & \multicolumn{2}{c}{\bf{Relevance}} &  \multicolumn{2}{c}{\bf{Latency (ms)}} & \multicolumn{2}{c}{\bf{Breakdown}}\\
S=SPLADE-HT1            & \%D & MRR@10 & R@1K &  MRT & P99 & I/O  & CPU  \\
        \hline
        S+Rerank & 0.01 & 0.421 & 0.980$^\dag$ & 158.7 & 217.3 & 125.2 & 33.5 \\
        S+LADR\textsubscript{fast} & 0.03 & 0.419 & 0.980$^\dag$ & 346.1 & 1013 & 1003 & 41.5 \\ 
        S+LADR\textsubscript{default} & 0.10 & 0.422 & 0.984 & 674.9 & 2849 & 2788 & 61.3 \\ 
        DiskANN & -- & 0.398$^\dag$ & 0.970$^\dag$ & 276.5 & 321 & 272 & 5.2    \\
        S+ DiskANN & -- & 0.417$^\dag$ & 0.984 &  307.7 & 352.2 &272& 36.4  \\ 
        SPANN & -- & 0.396$^\dag$ & 0.965$^\dag$ & 123.6 & 159.6  & -- & --    \\
        S+SPANN & -- & 0.420 & 0.989 &  154.8 & 190.8  &-- & -- \\
       $\blacktriangle$ S+CluSD & 0.05 & 0.425 & 0.986 & 55.6 & 94.4 & 17.5 & 38.1 \\
                \hline\hline
                \end{tabular}  }

        \label{tab:disktime}
           \vspace*{-5mm}
\end{table}
 
CluSD significantly outperforms all other methods by leveraging block I/O.
CluSD is 2.2x faster than the next fastest system (SPANN), while having a noticeably higher relevance.
DiskANN performs similarly to SPANN in terms of relevance, but is 4.97x slower than CluSD.
Reranking has a lower MRR@10 and recall@1k while being 2.85x slower.
Both variants of LADR have a lower MRR@10 than CluSD, but are 6.2x and 12.1x slower on average for LADR\textsubscript{fast} and LADR\textsubscript{default} respectively.
We find that  each  I/O operation has about a 0.15ms queueing and other software overhead on our tested PCIe SSD.
Thus, more fine-grained operations in reranking yields more overhead.
Proximity graph methods such as LADR rely on large amounts of fine-grained I/O operations and suffer significantly in terms of latency when embeddings are stored on disk.
In comparison, CluSD takes advantage of block I/O operations
because it fetches documents by clusters  and thus CluSD issues fewer I/O requests to search more documents in less time.

%% file: cikmexp_llama_ecir.tex
{\bf CluSD with LLaMA-2 based dense retrieval}.
Table~\ref{tab:RepLLaMA}
demonstrates that CluSD effectively supports selective RepLLaMA dense retrieval~\cite{ma2023finetuning} 
to take  advantage of   LLMs on CPUs. 
The ``Latency'' column includes sparse retrieval CPU time when ``S+'' is marked in the method name. 
S=SPLADE-HT1.
RepLLaMA  uses 145GB embedding space with 4K dimensionality.
All rows except the last row  conduct in-memory search assuming data fits into memory.
Row 1 shows  it takes 7.9 second CPU time to conduct in-memory full RepLLaMA  retrieval. 
Row 2  shows performance of RepLLaMA using OPQ quantization but without IVF selective search.
Row 3  shows performance of RepLLaMA using OPQ and  IVF selective search.
Use  of these optimization techniques leads to  a significant relevance degradation. 
As a reference, Row 4 lists the fusion of SPLADE and  full dense retrieval. 
CluSD visits  a small number of clusters from the 60,000 dense RepLLaMA clusters. 
CluSD  with SPLADE-HT1 takes about 39ms for in-memory search shown in  
Row 6, and  60ms with on-disk search shown in Row 8.  
RepLLaMA inference requires five Nvidia 32GB V100 GPUs  to host data and take 37ms GPU time. 
Thus, CluSD  can deliver 1 to 2 orders of magnitude of improvement in either total latency or infrastructure cost.
Rows 5 and 7 show the performance of CDFS, which is similar to CluSD for in-memory search, but 23.3\%
slower than  CluSD for on-disk search while achieving a similar relevance.

\comments{
\begin{verbatim}
MRR@10	R@1K	CPU Latency	Space
(1) RepLLaMA	0.412	0.990	7.9s	145GB
(2) RepLLaMA-ivfopq	0.365	0.915	97ms	6.1GB
(3) S + CluSD	0.424	0.986	60ms	149GB
(4) S + RepLLaMA	0.426	0.994	7.9s	149GB
(5) RepLLaMA-OPQ	0.384	0.990	666ms	2.4GB

\end{verbatim}
}

\begin{table}[htbp]
\vspace*{-1em}
\small
        \centering
        \caption{Selective fusion with  dense RepLLaMA and  sparse SPLADE} 
                \begin{tabular}{r r |l l|l l}
                        \hline\hline 
     &Method    & MRR@10 & R@1K & Latency & Space \\
                        \hline
1.  & RepLLaMA     & 0.412$^\dag$& 	0.990& 	7.9 s& 	145 GB\\
2.   &RepLLaMA-OPQ& 	0.384$^\dag$& 	0.990& 	666 ms& 	2.4 GB \\
3.  & RepLLaMA-IVFOPQ & 	0.365$^\dag$ & 	0.915$^\dag$ & 	97 ms & 	6.1 GB\\
4.   & S + RepLLaMA  & 	0.426& 	0.994& 	7.9 s& 	149 GB\\
5.   & S+CDFS & 0.425	& 0.987& 	41 ms& 	149 GB\\
6.   & S+CluSD & 0.424	& 0.986& 	39 ms& 	149 GB\\
7.   & S+CDFS (on-disk)& 	0.425	& 0.987& 	74 ms& 	149 GB \\
8.   & $\blacktriangle$ S+CluSD (on-disk)& 	0.424	& 0.986& 	60 ms& 	149 GB\\
                \hline\hline
                \end{tabular}  
        
 \vspace*{-5mm}
        \label{tab:RepLLaMA}
\end{table}

For the BEIR datasets,
RepLLaMA achieves an average NDCG of 0.561. 
While significantly faster than RepLLaMA, CluSD achieves an average NDCG of 0.541 while CDFS reaches 0.554.
However, CluSD's LSTM model for RepLLaMA was used in a zero-shot manner after being trained using SimLM on MS MARCO passage.
In the future we expect that further improvement in CluSD is possible if its LSTM is trained using RepLLaMA.

%% file: cikmexp_sparse_ecir.tex
\label{sect:evalsparse}
{\bf CluSD with other sparse models and quantization methods}.
Table~\ref{tab:sparse} examines CluSD's fusion of SimLM
with  other three sparse models.
Last column lists in-memory dense retrieval portion of latency.
For each sparse  model, CluSD boosts overall relevance.
As  CluSD's cluster selection  relies on  the overlap feature  of top sparse results with dense clusters,
the guidance from the fastest BM25 results is less accurate than other  sparse models. On the other hand, with a guidance from 
the slowest but stronger LexMAE retrieval, CluSD yields the highest MRR@10, comparable to full search. 

\comment{
\begin{table}[htbp]
	\centering
		\resizebox{\columnwidth}{!}{%
		\begin{tabular}{ r r r r r  r r r}
			\hline\hline
			& \multicolumn{2}{c}{MSMARCO Dev}& \multicolumn{2}{c}{DL19}& 
\multicolumn{2}{c}{DL20} \\
			& \% n& MRR@10 & R@1K & NDCG@10 & R@1K &  NDCG@10 & R@1K\\
   			\hline
 D& -- & 0.416& 0.988& 0.720& 0.744& 0.703&0.755\\
            $\blacktriangle$  S $+$ D & -- & 0.425 & 0.988 &  0.740& 0.816&   0.731& 0.819\\
            \bf S $+$ \bf{CluSD} & 0.5 &  0.426 & 0.987 &  0.744&  0.820& 0.734&  0.818\\
            \hline
            \multicolumn{8}{c}{OPQ M=128 setting, Storage 1.1GB, Report S+*}\\
            \hline
            IVFOPQ &10 & 0.404 & 0.987 & 0.713 & 0.818 & 0.722 & 0.817 \\
            & 5 & 0.394 & 0.987 & 0.687 & 0.817 & 0.706 & 0.817 \\
            & 2 & 0.374 & 0.986 & 0.656 & 0.815 & 0.700 & 0.817 \\
            HNSWOPQ & & & & & & \\
            \bf{CS} & 0.5 & 0.417& 0.986& 0.742 & 0.818 & 0.735  &0.815 \\
            \hline
            \multicolumn{8}{c}{OPQ M=64, Storage 0.6 GB, Report S+*}\\
            \hline
            IVFOPQ & 10 & 0.393 & 0.986 & 0.676 & 0.814 & 0.730 & 0.818 \\
            & 5 & 0.384 & 0.985 & 0.659 & 0.814 & 0.717 & 0.817 \\
            & 2 & 0.368 & 0.985 & 0.643 & 0.814 & 0.707 & 0.816 \\
            HNSWOPQ & & & & & & & \\
            \bf{CS} & 0.5  & 0.403& 0.987& 0.729 & 0.812 &0.724 & 0.814 \\           
			\hline\hline
		\end{tabular}
		}
	\caption{RetroMAE
}
	\label{tab:retromae}
\end{table}
}

 \vspace*{-5mm}
\begin{table}[htbp]
	\centering
    \caption{ CluSD with  LexMAE, uniCOIL \& BM25  sparse models}
		\resizebox{0.75\columnwidth}{!}{%
		\begin{tabular}{ r |l l  |ll |ll |r }
			\hline\hline
                Sparse model& \multicolumn{2}{c|}{\bf{LexMAE}} & \multicolumn{2}{c|}{\bf{uniCOIL}}   & \multicolumn{2}{c|}{\bf{BM25T5}} & \bf{Dense}\\ 
\cline{1-7}
			& {MRR} & {R@1K}  & {MRR} & {R@1K}   & {MRR} & R@1K  & \bf{latency}\\
   \hline
   \multicolumn{8}{c}{\bf{Uncompressed SimLM  embeddings  with 27.2 GB space}}\\
			\hline
			S & 0.425& 0.988& 0.347$^\dag$ & 0.957$^\dag$& 0.259$^\dag$ &  0.935$^\dag$& -\\
                $\blacktriangle$  S $+$ D &    0.430& 0.990& 0.413 & 0.985& 0.415& 0.985& 1674ms\\
                S $+$ Rerank& 0.432& 0.988& 0.415& 0.957$^\dag$& 0.409& 0.935$^\dag$& 21.9ms\\
                {\textbf{S $+$ CluSD}} & 0.431& 0.989& 0.414 &  0.980&  0.410&  0.980& 12.1ms\\
                \hline
                \multicolumn{8}{c}{\bf{OPQ m=128. SimLM embedding space  1.1 GB}}\\
\hline
                S $+$ D-IVFOPQ & 0.383$^\dag$& 0.988& 0.354$^\dag$& 0.961$^\dag$& 0.369$^\dag$& 0.975$^\dag$& 21.9ms\\
                \bf{S $+$ CluSD} & 0.429& 0.989& 0.404 & 0.980&  0.402$^\dag$& 0.975$^\dag$& 9.8ms\\
            \hline\hline
		\end{tabular}
		}
	
\vspace*{-6mm}
	\label{tab:sparse}
\end{table}

%% file: cikmexp_compress_ecir.tex

{\bf CluSD with advanced quantization models}.
Table~\ref{tab:compress} shows the use of two other recent quantization methods 
DistillVQ~\cite{Xiao2022Distill-VQ} and JPQ~\cite{2021CIKM-JPQ-Zhan} 
with in-memory CluSD for MS MARCO Passage Dev set. 
For JPQ, we use its released model and compression checkpoints and follow its default setting with $m=96$. 
For DistillVQ, we use its released code to train compression on the RetroMAE model using its default  value $m=128$.
While switching to a different quantization affects compression and  relevance, 
this result shows that CluSD is still effective and outperforms other baselines.

\comments{
\begin{verbatim}
DistillVQ	MRR@10	Recall@1K	Latency (ms)
IVF	0.365	0.899	20.8
CluSD	0.393	0.977	9.7
SPLADE+ IVF	0.392	0.987	52.0
SPLADE+CluSD	0.417	0.987	40.9

JPQ	MRR@10	Recall@1K	Latency (ms)
IVF	0.326	0.917	21.3
CluSD	0.341	0.969	11.2
SPLADE+IVF	0.379	0.985	52.5
SPLADE+CluSD	0.392	0.985	42.4
\end{verbatim}
}

\begin{table}[htbp]
        \centering
        \caption{ Use of CluSD with   DistillVQ and JPQ quantization}
                \begin{tabular}{ r |l l l |ll l}
                        \hline\hline
                Quantization & \multicolumn{3}{c|}{\bf{DistillVQ}} & \multicolumn{3}{c}{\bf{JPQ}}   \\
                        \hline
S=SPLADE-HT1                        & {MRR@10} & {R@1K}  & {Latency } & {MRR@10} & {R@1K}    & {Latency}\\
                        \hline
                        IVF (2\%)  & 0.365$^\dag$ &	0.899$^\dag$ &	20.8 ms
& 0.326$^\dag$	&0.917$^\dag$ &	21.3 ms\\
S+ IVF (2\%)	&0.392$^\dag$ &	0.987&	52.0 ms
& 0.379$^\dag$ & 	0.985& 	52.5 ms\\
$\blacktriangle$ S+CluSD&	 0.417 &	0.987 &	40.9 ms
 &0.392 &	0.985 &	42.4 ms\\
\hline           
\hline           
                \end{tabular}%
\vspace*{-5mm}  
        \label{tab:compress}
\end{table}

%% file: cikmexp_feature_ecir.tex
{\bf Design options for CluSD}.
\label{sect:evalLSTM}
Table~\ref{tab:cluster_feat} investigates the impact of the following design options 
for cluster selection in CluSD with  the MS MARCO Dev set. 
For Stage I selection, we explore two options:
1) SortByDist: Choose top clusters sorted by the distance of their  centroids to  the  query embedding.
2)  SortByOverlap:  Choose top  clusters sorted  by $P(C_i, B_j)$ features discussed in Section~\ref{sec:clusd}.
Namely, use the overlap degree of clusters with top result bins of sparse retrieval. 
For Stage II selection, we explore three options:
    1) XGBoost: Use the boosting tree XGBoost model to detect if a cluster should be visited.
The features used are the same ones discussed in Section~\ref{sec:clusd}. This is a pointwise approach, as we view query-cluster pairs as independent samples, and the decision on one cluster does not affect other clusters for the same query.
    2) RNN: Use a vanilla RNN model. The feature and sequence setup is the same as our LSTM model. 
    3) LSTM: the model we used in CluSD.
 We also assess  the impact  of removing a  feature group used in CluSD:
1) w/o inter-cluster dist: removing  the inter-cluster distance feature group;
2) w/o S-C overlap: removing  the overlap-degree feature of top sparse results with each cluster.



\begin{table}
\caption{Design options for  CluSD. $\blacktriangle$ is the default choice.}
\centering
\resizebox{0.8\columnwidth}{!}{%
  \begin{tabular}{ r |r r| r r| r }
			\hline\hline
			 Avg \#clusters targeted  & \multicolumn{2}{c}{3} & \multicolumn{2}{c}{5} & Time \\
			& MRR@10 & R@1K& MRR@10 & R@1K & ms\\
                   \hline
                 \multicolumn{6}{l}{\it{Stage II is removed. Stage I options only}}\\
                 SortByDist & 0.297 & 0.655 & 0.331 & 0.740  & 0 \\
                $\blacktriangle$ SortByOverlap & 0.384 & 0.867 & 0.401 & 0.917 & 0.2 \\
                   \hline
                \multicolumn{6}{l}{\it{Stage I=SortByDist;  Stage II model options}}\\
                 XGBoost& 0.398 & 0.888 & 0.404 & 0.929 & 192 \\
                 RNN & 0.404 & 0.908 & 0.406 & 0.938 & 2.8 \\
                 LSTM & 0.406 & 0.923 & 0.406 &0.943 & 2.8\\
                   \hline
             \multicolumn{5}{l}{\it{ Stage II=SortByOverlap;  Stage II LSTM feature group options}}\\
             w/o inter-cluster  dist& 0.402 & 0.915 & 0.406 & 0.941 & --\\
             w/o  S-C overlap& 0.289 & 0.638 & 0.325 & 0.730 & --\\
             \hline 
             $\blacktriangle$  Default   &  0.408 & 0.943 & 0.410 & 0.953 & 2.8 \\
              \hline\hline
		\end{tabular}
}
        \label{tab:cluster_feat}
\vspace{-5mm}
\end{table}

The last row marked ``Default'' is the final version used by CluSD with SortByOverlap and LSTM.
To have a fair comparison, this table assumes on average either 3 or 5 clusters are targeted by CluSD.
For XGBoost, RNN or LSTM, we set their prediction threshold so that 
the average number of clusters is close to the targeted number.
The last column of this table is the time spent to select clusters. 
Table~\ref{tab:cluster_feat} shows that  the RNN or LSTM based prediction achieves  higher MRR@10 and recall compared to 
XGBoost. 
Feature exploration indicates that the overlap degree is important for  cluster selection
while the inter-cluster distance feature group is also useful.

\begin{figure}[h!]
    \centering
    \includegraphics[scale=0.18]{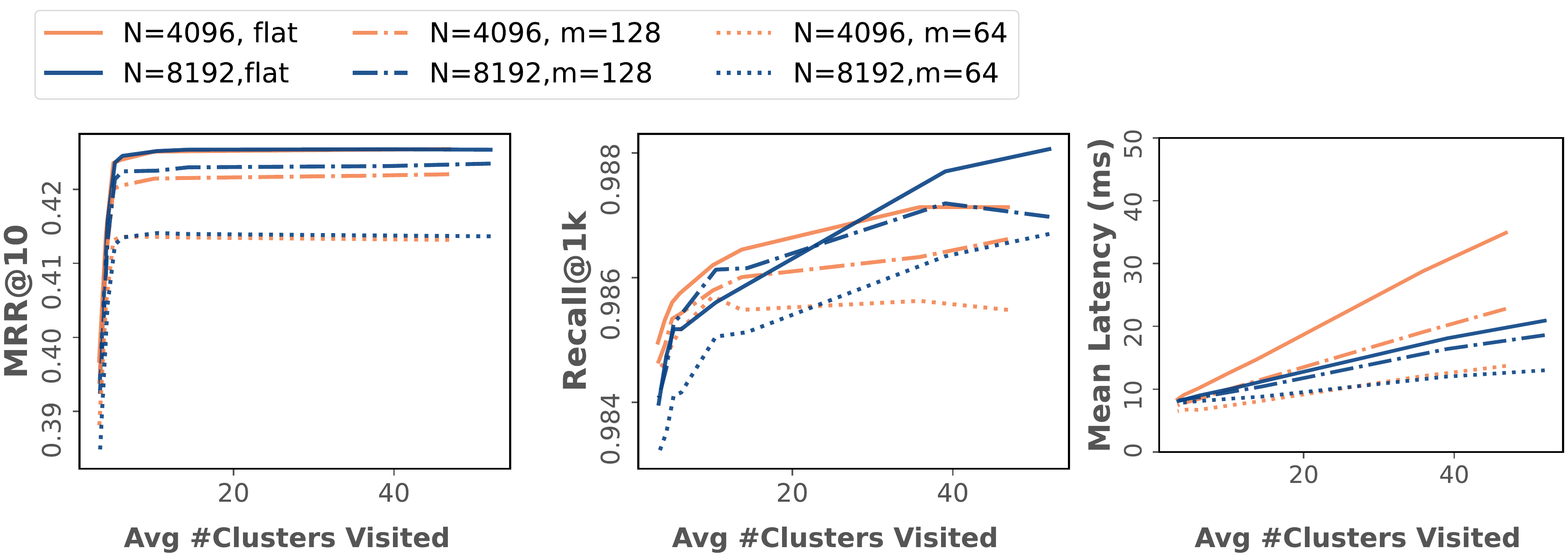}
    \caption{CluSD relevance and latency vs.  the average  number of clusters selected }
    \label{fig:ncluster}
 \vspace*{-4mm}
\end{figure}

\textbf{Impact of  cluster partitioning size}.
CluSD pre-partitions the embeddings into $N$ clusters. 
Figure~\ref{fig:ncluster} shows the impact when $N$ is  4096 and  8192 for MS MARCO Dev.
Stage I of CluSD uses  $n=128$ in this figure. 
A  solid line refers to the use of uncompressed dense embeddings while  a dash-dot line refers to OPQ quantization with m=128, 
and a dotted line refers to OPQ quantization  with m=64. The x-axis is the average number of clusters selected by CluSD.
We vary the selection threshold parameter $\Theta$ in CluSD to achieve the different 
average number of clusters selected. 
With both $N$ values,
MRR@10 exceeds 0.420 when more than 10 clusters are selected.
As more clusters are selected,  the recall ratio increases. 
Different $N$ values impact latency. 
For example, when 10 clusters are selected on average, 
CluSD  scans through 0.1\% and  0.2\% of embeddings, corresponding to N=4096 and 8192, respectively. 
When $N= 8192$, its latency is reduced
because CluSD time complexity is proportional  to  the size per cluster and the number of clusters selected.
 \vspace*{-5mm}

%% file: ecirconcl.tex
\vspace*{-4mm}
\section{Concluding Remarks}

CluSD is a lightweight cluster-based partial dense retrieval with
fast CPU response time and competitive relevance.
Compared to CDFS~\cite{2024SIGIR-CDFS-Yang},
CluSD does not make any strong statistical assumptions about the sparse ranking result distribution.
The evaluation results show that  CluSD can effectively select top dense
clusters with a performance  reasonably competitive to CDFS in relevance and latency
even CluSD incurs small LSTM computing overhead.  
More studies and tuning could be conducted
in the future in evaluating CluSD and CDFS.
     

The evaluation shows CluSD significantly outperforms cluster-based partial dense retrieval with IVF in relevance.
Under the same time budget, CluSD with negligible extra space overhead 
delivers better or similar relevance than HNSW and LADR that rely upon
a document-level proximity graph.
Further, CluSD is up-to 12.1x faster than them
when embeddings 
do not fit in memory.
When compared to on-disk ANN methods DiskANN and SPANN~\cite{NEURIPS2019_DiskANN,2023Web-Filtered-DiskANN},
CluSD is 2.2x and 4.97x faster respectively, while achieving better relevance
by leveraging sparse retrieval.
Compared to graph navigation approaches, 
CluSD does not need  a  document-level proximity graph,  and  it conducts faster block I/O when fetching clustered document embeddings  
from disk.  

\comments{
CluSD is also stronger and more versatile than reranking. 
Simple reranking does surprisingly well if sparse results are strong. But as stated in~\cite{2022CIKM-MacAvaneyGraphReRank,2023SIGIR-LADR},
when the sparse retriever doesn’t achieve high relevance, reranking is too 
restrictive and recall@1K does not increase as shown in Table 5. 
CluSD’s advantage becomes more significant in zero-shot retrieval as shown in Table 6 with 13 BEIR datasets. CluSD’s average NDCG number is 
7.2\% higher than reranking with RetroMAE. With SimLM, CluSD is 4\% higher than reranking.
In addition, when embeddings are not available in memory, reranking is 2.85x slower in mean latency.

CluSD works well  with LLaMA-2 based RepLLaMA dense  model for the tested datasets.
More such models  can be developed in the future using large language models of high dimensionality,
and our study sheds a light on how to take advantages of such dense models
on low-cost CPU servers, especially when their embeddings cannot fit into memory and are hosted on disk.

}
\comments{
Our evaluation has not used ColBERT with a multi-vector representation because 
ColBERTv2's MRR@10~\cite{Santhanam2021ColBERTv2}  for MS MARCO is 0.397, which is lower than that the single-vector dense models studied in this paper. 
CluSD techniques are  orthogonal to ColBERTv2 and could be used with ColBERTv2 after small modifications.
}

{\bf Acknowledgments}.
We thank anonymous referees for their valuable comments.
This work is supported in part by U.S. NSF IIS-2225942  and has used the computing resource of the ACCESS program supported by NSF.
Any opinions, findings, conclusions or recommendations expressed in this material
are those of the authors and do not necessarily reflect the views of the U.S. NSF.